# HUI-Audio-Corpus-German:
# A high quality TTS dataset

Pascal Puchtler, Johannes Wirth and René Peinl

Hof University of Applied Sciences, Alfons-Goppel-Platz 1, 95028 Hof, Germany
`Pascal.Puchtler@iisys.de`
`Johannes.Wirth@iisys.de`
`Rene.Peinl@iisys.de`

**Abstract.** The increasing availability of audio data on the internet lead to a multitude of datasets for development and training of text to speech applications, based on neural networks. Highly differing quality of voice, low sampling rates, lack of text normalization and disadvantageous alignment of audio samples to corresponding transcript sentences still limit the performance of deep neural networks trained on this task. Additionally, data resources in languages like German are still very limited. We introduce the "HUI-Audio-Corpus-German", a large, open-source dataset for TTS engines, created with a processing pipeline, which produces high quality audio to transcription alignments and decreases manual effort needed for creation.

## 1   Introduction

Performance of text to speech (TTS) systems has increased vastly over the past decade, primarily by leveraging deep neural networks (DNNs) [1], which in turn lead to higher acceptance by end-users [1, 2]. TTS with DNNs is a two-stage process with a Mel-Spectrogram as an intermediate output that a vocoder converts into the final audio file. Tacotron 2, is one of the most popular models for the first stage and achieved a mean opinion score (MOS) of 4.53 on a five point scale for the English language [3] using a modified WaveNet [4] as vocoder (second stage). Globally operating companies have started to incorporate TTS engines for human machine interaction into their products, like home assistants, cars or smartphones [5]. To achieve such good results, training data must be available in large enough quantity and high enough quality. For English language, there are high quality datasets like LJ speech [6] and LibriTTS [7], which are commonly used and produce good results [8, 9].

In languages other than English, high quality training data is scarce and creation of new datasets often require unfeasibly high efforts as well as time. This is especially true for researchers in the domain of audio processing as well as smaller businesses so they mostly have to resort to freely available data, in order to utilize this technology.

In this paper we introduce a new, open-source dataset for TTS, called **HUI**-Audio-Corpus-German (**H**of **U**niversity – **I**nstitute for information systems) for the German language, which consists of over 326 hours of audio snippets with matching transcripts, gathered from librivox.org and processed in a fine-grained refinement pipeline. The dataset consists of five speakers with 32 – 96 hours of audio each to construct single speaker TTS models, as well as 97 hours of audio from additional 117 speakers for diversity in a multi-speaker TTS model. For every speaker, a clean version with high signal-noise distance has been generated additionally, further increasing quality. The underlying goal was to create a German dataset with the quality of LJ Speech [6] that is more comprehensive than the one from M-AILABS [10]. The dataset[1] as well as the source code[2] are open source and freely available.

The remainder of this article is organized as follows. We start discussing related work on freely available datasets in English and German and derive requirements for an own dataset from it in section 2. We introduce the data processing pipeline that we used to create our own dataset in section 3. We present our own dataset in section 4 and discuss its advantages over existing datasets, before concluding the article with a summary and outlook.

## 2 Related Work

LJ Speech is a well-known audio dataset, that achieves good results in state-of-the-art TTS models [11]. This public domain dataset comprises almost 24 h of speech recordings by a single female speaker reading passages from seven non-fiction books. In total, there are 13,100 utterances with an average length of 6.6 s [6]. It is used by many TTS research papers, although the recordings exhibit a certain degree of room reverberation.

**Table 1.** Dataset overview

| Corpus | License | Duration (hours) | Sampling rate (kHz) | Total speakers |
|---|---|---|---|---|
| Thorsten-Voice neutral [12] | CC0 | 23 | 22.05 | 1 |
| CCS10 – German [13] | CC0 | 17 | 22.05 | 1 |
| M-AILABS – German [10] | BSD | 237 | 16 | 5+[3] |
| MLS – German [14] | CC By 4.0 | 3,287 | 16 | 244 |
| **HUI-Audio-Corpus-German** | CC0 | 326 | 44.1 | 122 |

In contrast to this single-speaker dataset, LibriTTS is a popular dataset that can be used for multi-speaker training [7]. The corpus consists of 585 hours of speech data at 24kHz sampling rate from 2,456 speakers and the corresponding texts. It is derived from the LibriSpeech corpus [15], which is tailored for automatic speech recognition (ASR), but comes with a number of problems regarding its use for TTS. These are a low sample rate (16 kHz), removed punctuation and varying degrees of background noise [7]. While a selection of TTS-ready datasets already exists in German (see Table 1), most of them have similar quality issues, which is in turn reflected in output quality of trained models.

---

[1] https://opendata.iisys.de/datasets.html#hui-audio-corpus-german
[2] https://github.com/iisys-hof/HUI-Audio-Corpus-German
[3] The data set consists of 5 named speakers plus others aggregated in mixed.

According to our demands, a TTS dataset of high quality should therefore fulfil at least the following requirements:

1. A minimum recording duration of 20 hours per speaker (for single speaker dataset)
2. Audio recordings with sampling rates of at least 22,050 Hz (as suggested by [7])
3. Normalization of text (resolution of abbreviations, numbers etc., see [7])
4. Normalization of audio loudness
5. Average audio length between 5 to 10 seconds (inspired by [6], with ø 6.6 s)
6. Inclusion of pronunciation-relevant punctuation
7. Optional: Preservation of capitalization (as suggested by [7])

Thorsten-Voice neutral [12] is the only dataset, which meets all our requirements, with 23 hours of audio from a single speaker in good to medium quality and 22.05 kHz. However, it is read nearly over-emphasized leading to unnaturally sounding TTS results in our experiments. CSS10 - German [13] is a collection of single speaker speech datasets in ten languages. Its German part has a good text normalization as well as a sufficient sampling rate. However, the amount of data is quite low with not even 17 hours of German speech from a single female speaker (Hokuspokus). M-AILABS [10] have compiled speech data from five main speakers with 19, 24, 29, 40 and 68 hours of speech. It contains mostly perfect text normalization, but the sampling rate of the recordings is only 16 kHz. Multi-lingual LibriSpeech (MLS) is an automatically generated dataset for multiple languages [14]. In the German variant with a massive 3,287 hours of audio, however, errors have occurred in the normalization of the texts. Numbers are e.g., completely missing in the text. Additionally, the sampling rate is not sufficient. All these datasets, except Thorsten-Voice neutral, are derived from LibriVox.

## 3    Data Processing Pipeline

To create a high-quality TTS dataset fulfilling the previously described requirements, a fine-grained pre-processing pipeline was constructed, which generates audio-transcript pairs, featuring automated download of data, very precise alignment of audio files and transcripts with utilization of a deep neural network, audio/text normalization and further processing (see **Fig. 1**).

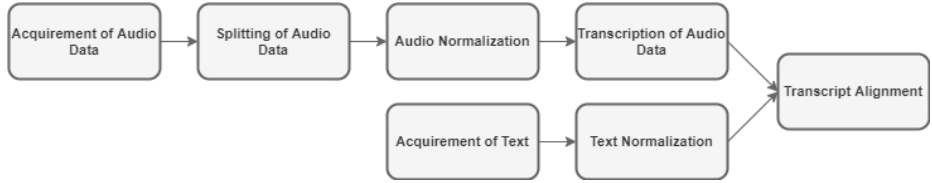

**Fig. 1.** Data processing pipeline overview.

### 3.1 Acquirement of Suitable Audio Data

We use LibriVox, a web platform, offering "free public domain audiobooks"[4] in several languages, as a source for audio. The available audio files are read and created by volunteers in various lengths and recording qualities. All authors of the texts read aloud either passed away more than 70 years ago or their publishers agreed to a free publication. Thus, neither books nor recordings are subject to copyright claims. The LibriVox API[5] presents a convenient way to retrieve metadata about available audiobooks and is leveraged to create an automated download process for audiobooks. Audio files are available in different sampling rates on the platform. For data generation, only files with sampling rates of 44.1 kHz were considered, as generally higher sampling rates allow for greater flexibility in terms of further sampling rate adjustment.

### 3.2 Splitting of Audio Data

For training with neural networks, short audio snippets with lengths ranging from 10 to 20 seconds are preferred in comparable works such as [14]. Shorter recordings result in a worse sentence melody for longer inference outputs after training. Longer input data lead to slower loss convergences at training time and thereby to higher computational complexity. However, other successful TTS datasets like LJspeech have a range of 1-10 seconds of audio [6]. Therefore, the thresholds of recording duration were set to a range of 5 to 40 seconds in order to preserve emphasis on full sentences, mainly beginning and ending.

Recordings are split into the range of desired lengths by a search for silent audio segments of at least 0.2 seconds. Since the silence for each audio file varies considerably, it is not possible to define a fixed volume for silence. This problem is circumvented by an increase of the Decibel (dB) value for silence until the longest audio snippet from a recording is shorter than the maximum length threshold. Afterwards, audio snippets shorter than the minimum are combined with preceding or following snippets until the total length of those exceed the minimum length value. Recordings are split at the centre of a silent section.

---

[4] https://librivox.org/
[5] https://librivox.org/api/info

### 3.3 Audio Normalization

Audio normalization is hereinafter defined as adjustment of the volume of audio files to a uniform value. Experiments suggest that -20 dB is considered useful for filtering background noise. Moreover, the loudness of the majority of data we acquired already had a level close to this threshold. This is supported by the use of -24dB in the Thorsten-Voice dataset [12]. For the implementation pyloudnorm[6] is used. Additionally, a fade in/out of 0.1 seconds is applied to the beginning and end of recordings to further filter out undesired sounds such as breathing.

### 3.4 Transcription of Audio Data for Subsequent Alignment

Each audio file is transcribed. A trained Deep Speech model from [17] in the most recent version[7] is utilized. Inferences are created in conjunction with a 3-gram KenLM language model, which is as provided together with the Deep Speech model. The model uses 16kHz as input sample rate. While the base model had achieved a word error rate (WER) of 21.5% with [17], the used version (based on a newer Deep Speech implementation) was trained with additional datasets, but no new benchmark data was published.

### 3.5 Acquirement of Text for Audio Data

For each audio book, LibriVox provides a link to the original text. For German, these are mainly hosted on projekt-guttenberg.de[8] and guttenberg.org[9], offering public domain books and literary prose works. Our solution downloads the texts automatically and parses them for further processing.

### 3.6 Text Normalization

Preparation of transcripts partially has to be conducted in manual processes, due to individual differences of the speakers. An overview of the replacements used can be found in **Table 2**.
**Numbers.** In German, the correct normalized form of ordinal numbers depends on grammatical gender, case as well as grammatical number. This increases the number of possible normalizations by a large factor at each occurrence, compared to e.g., English.
**Abbreviations**. Partially, abbreviations written in the same way have to be mapped to different normalized words, which significantly complicates automation.
**Censorship.** Parts of the texts were censored, e.g., because of German history, mainly terms and names from the national socialists era. Different speakers dealt with this kind of symbol sequences in various manners, which in turn leads to the need of manually comparing recordings with transcripts in order to gain best possible audio to transcript alignments.

---

[6] https://github.com/csteinmetz1/pyloudnorm
[7] https://github.com/AASHISHAG/deepspeech-german#trained-models
[8] https://www.projekt-gutenberg.org/
[9] https://gutenberg.org

**Footnotes.** Some of the texts contain footnotes. These are again treated differently by speakers. The most common ways are 1) omit completely 2) read the number and read the footnote at the end of the page 3) omit the number and read the footnote immediately. Additionally, in some cases the word "Fussnote" (footnote) is added by a speaker. In other cases, the word is explicitly written in the text.

**Comments.** The texts partly contain comments in round brackets. These are only partly read out loud. Depending on the speaker, "Kommentar Anfang" (comment beginning) and "Kommentar Ende" (comment end) are added.

Table 2. Replacements for text editing.

| Original text | Normalized text | Type of normalization |
| --- | --- | --- |
| XIII | Siebzehn | roman numeral |
| III. | der dritte | roman ordinal number nominative case |
| 51,197 | einundfünfzig komma eins neun sieben | decimal number |
| 5½ | fünf einhalb | numbers with fractures |
| 30. | Dreißigsten | ordinal number dative case |
| 1793 | Siebzehnhundertdreiundneunzig | year |
| 1804/05 | achtzehnhundertvier fünf | range of years |
| 1885/86 | achtzehnhundertfünfundachtzig bis sechsundachtzig | range of years |
| 50 000 | Fünfzigtausend | decimal number without separator |
| 4,40 Mk. | Vier Mark vierzig | sum of money (in specific currencies) |
| E.Th.A. Hoffmann | Ernst Theodor Amadeus Hoffmann | name complete |
| Prof. Dr. Sigm. Freud LL. D | Professor Doktor Sigmund Freud Doktor of Law | name complete |
| Pf...sche | Pfffsche | emphasis in the text |
| *** | Punkt Punkt Punk | censorship pronounced |
| St. | Sankt | abbreviation |
| a. D. | a D | abbreviation |
| = | Ist | abbreviation |

### 3.7 Transcript Alignment

At this step, the original normalized text as well as artificially generated transcripts are present. These are needed to create the best possible automated alignment between read aloud words from audio snippets and the corresponding transcripts. The generated transcripts mostly follow the same order as the original text. Note that spoken intro and outro sequences do not have corresponding transcripts.

In the following step, a positional alignment between original and artificially generated transcript sentences is to be achieved. The original text is a long string without alignment to the recordings. However, it contains punctuation, capitalization and error-free words. The transcripts of the recordings are a list of texts with assignment to a recording. However, they are partly incorrect in text and without punctuation and capitalization, because of the error rate of the German Deep Speech [17].

The first transcript is near the beginning of the original text, with means that the alignment search area can be clearly limited. In this search area, the distance between each possible subarea and the transcript is formed. The range with the smallest distance is called match and is kept as the ground truth alignment between the original text and the recording. As distance $d(s1, s2)$ with the strings $s_1$ and $s_2$ we use a modified version of the Levenshtein Distance [18]:

$$d(s1, s2) \stackrel{\text{def}}{=} \frac{Levenshtein-Distance(s_1, s_2)}{\max(length(s_1), length(s_1))} \quad (1)$$

Next the search area is moved by the length of the match, there the search for the next match is repeated.

Now we have an associated part of the original text for each generated snippet. Due to various possible problems, the quality of the hit may not be sufficient. This can be determined by the distance. By testing we have found that hits above a value of 0.2 should be discarded.

The transitions from one text snippet to the next are always a problem, as words can appear twice or not at all. To overcome this problem, a transition is called perfect if both matches are exactly adjacent to each other.

A text snippet is of sufficient quality for us, if all the following conditions are met:

- The text snippet has a distance of less than 0.2
- The preceding and following text snippet has a distance of less than 0.2
- The transition to the previous and following text snippet is perfect

This way we can be sure that all text snippets are assigned in the best possible way and that our final data set has as few errors as possible. However, this also means that some of the text snippets are discarded.

## 4 Dataset Summary

### 4.1 Full Dataset

The dataset was statistically evaluated for each included speaker (see Table 3). The following aspects are considered:

**Speakers.** Number of Speakers.
**Hours.** Total audio data length in hours.
**Count.** Count of audio-transcript pairs.
**MVA.** In each audio snippet, the frame with the minimum volume (in dB) is determined. An average is calculated over the minimum volume values in all audio snippets. This is defined as Minimum Volume Average (MVA). The standard deviation is indicated in parentheses.
**SPA.** Each audio snippet can be divided into silence and speech, through RMS. The proportion of silence is measured for each audio snippet and an average is formed over the dataset. This metric is defined as silence proportion average (SPA). The value in parentheses represents the standard deviation of the data.
**UW@1.** Count of all unique words that occur in the transcripts. UW@1 describes the diversity of the transcripts. A larger value is an indication for higher coverage of the German vocabulary.
**UW@5.** Count of all unique words that occur at least five times in the transcripts. Extension of the UW@1 metric. The higher the frequency of unique words within the dataset, the less impact one-time poorly pronounced words have on the training process of TTS models.

**Table 3.** Subset overview - full

| Subset | Speakers | Hours | Count | MVA | SPA | UW@1 | UW@5 |
|---|---|---|---|---|---|---|---|
| Bernd Ungerer ♂ | 1 | 97 | 35k | -60 (6.1) | 20 (6.5) | 33.5k | 9.1k |
| Hokuspokus ♀ | 1 | 43 | 19k | -45 (14.3) | 18 (10.4) | 33.7k | 5.9k |
| Friedrich ♂ | 1 | 32 | 15k | -52 (8.9) | 27 (9.6) | 26.6k | 5.0k |
| Karlsson ♂ | 1 | 30 | 11k | -60 (4.4) | 20 (7.0) | 26.4k | 4.5k |
| Eva K ♀ | 1 | 29 | 11k | -56 (4.9) | 18 (7.6) | 23.2k | 4.4k |
| Other ♂/♀ | 117 | 96 | 38k | -55 (14.0) | 20 (9.5) | 60.9k | 11.7k |
| Total | 122 | 326 | 130k | -55 (11.5) | 20 (8.9) | 105k | 25.4k |

**Table 4.** Subset overview - clean

| Subset | Speakers | Hours | Count | MVA | SPA | UW@1 | UW@5 |
|---|---|---|---|---|---|---|---|
| Bernd Ungerer ♂ | 1 | 92 | 33k | -61 (5.6) | 21 (6.0) | 31.8k | 8.8k |
| Hokuspokus ♀ | 1 | 27 | 11k | -57 (2.8) | 22 (6.4) | 24.4k | 4.1k |
| Friedrich ♂ | 1 | 21 | 9.6k | -56 (5.8) | 26 (7.6) | 21.1k | 3.6k |
| Karlsson ♂ | 1 | 29 | 11k | -60 (3.7) | 21 (6.4) | 25.4k | 4.3k |
| Eva K ♀ | 1 | 22 | 8.5k | -57 (4.0) | 19 (6.4) | 18.8k | 3.4k |
| Other ♂/♀ | 113 | 64 | 24k | -63 (9.1) | 22 (7.0) | 48.4k | 8.6k |
| Total | 118 | 253 | 97k | -60 (6.6) | 21 (6.8) | 87.9k | 21.0k |

### 4.2 Clean Dataset

The audio quality of the individual audio snippets may well vary, caused by e.g. background noise or poor recording quality. For this reason, a clean variant was created for each subset. Using thresholds for minimum volume and silence proportion, each dataset was filtered, the resulting datasets are considered "clean" sets. The following thresholds were used:

$$min\ volume < -50\ dB\ \wedge\ 10\% < silence\ proportion < 45\% \qquad (2)$$

The minimum volume can be seen as a simplified version of the signal-noise-ratio, since in silent parts, only background noise is generating sound. A statistical evaluation of the resulting clean variants is presented in **Table 4**.

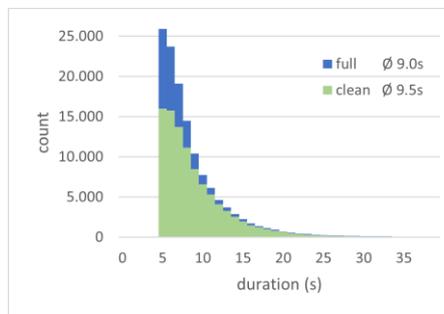
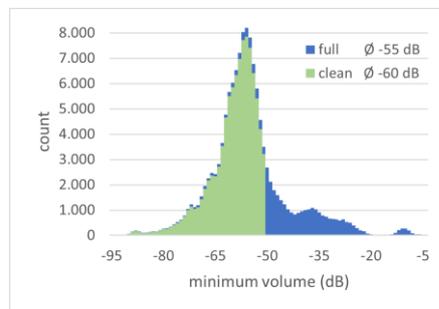

**Fig. 2.** Histogram of the audio duration for all speakers.

**Fig. 3.** Histogram of minimum volume for all speakers.

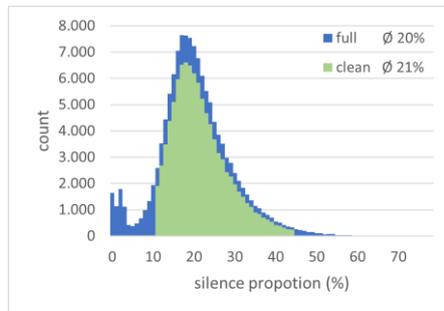
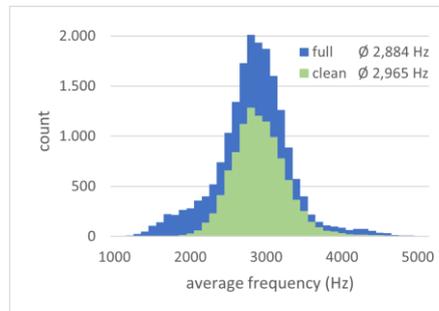

**Fig. 4.** Histogram of the silence proportion for all speakers.

**Fig. 5.** Histogram of the average frequency for the speaker "Hokuspokus".

Figs. 2 to 4 show histograms of the full and clean datasets for all speakers. Figure 2 shows the distribution of durations for all audio snippets. A strong tendency towards the 5-10 second range as well as the exclusion of any snippets under the length of 5 seconds can be observed. Furthermore, there are no audio snippets under 5 seconds. Fig. 3 demonstrates a large variance with respect to minimum volume threshold, which is significantly lower within the clean dataset in comparison. Fig. 4 shows the proportion of silence within the audio snippets. A concentration at 0% and no values above 70% can be detected. Furthermore, boundary values are recognizable for the clean dataset. Fig. 5 depicts the average speech frequency of the audio snippets for speaker Hokuspokus. It shows that our normalization reduced the standard deviation of frequencies per speaker significantly.

Considering the described figures, it can be hypothesized that training a TTS model using the clean datasets will lead to a potentially better result, since audio snippets contained in the clean dataset show a higher coherence, primarily in terms of frequency spectrum, proportion of silence and minimum loudness. The duration of audio snippets is insignificantly higher within the clean dataset (Ø 9.5s), compared to the full dataset (Ø 9.0s).

### 4.3 Discussion

The generated HUI-Audio-Corpus-German is compared to the previously established requirements for a state-of-the-art TTS dataset.

**1) Minimum duration of 20 hours per speaker.** For the five main speakers, this goal is achieved. In addition, the "other" subset consists of several speakers, none of which has exceeded the set threshold of 20 hours.
**2) Sampling rate of at least 22,050 Hz.** Each audio snippet in the HUI-Audio-Corpus-German has a sampling rate of 44.1 kHz.
**3) Normalization of text.** An automated check for digits, abbreviations and special characters as well as a thorough manual analysis of transcript samples confirmed the required grade of text normalization.
**4) Normalization of audio loudness.** All audio snippets are normalized according to the requirement.
**5) Average audio length of 5 to 10 seconds.** The full dataset as an average audio length of 9.0 seconds and the clean dataset of 9.5 seconds, thus averages of both sets are within the specified limits.
**6) Inclusion of pronunciation-relevant punctuation.** As punctuation relevant symbols, period (.), question mark (?) exclamation point (!), comma (,) and colon (:) were chosen. All other punctuations were either transformed or completely removed.
**7) Preservation of capitalization.** Capitalization of transcripts is preserved by default. The statistics in **Table 3** and **Table 4** show, that even the longest single speaker dataset contains only 33.5k unique words, compared to 105k for the whole dataset. This is due to the focus of the books that were read and can be an issue for open domain TTS.

### 4.4 Evaluation with Tacotron 2

In order to verify and compare the overall quality of full and clean datasets as well as their effects on convergence of loss in a deep neural network for TTS, both variations of subsets by the speaker "Hokuspokus" were selected to be used for the training of multiple Tacotron 2 [19] models in conjunction with a Multi-band MelGAN [2] as vocoder. For comparability, all models were trained with identical configurations.

While training loss (Fig. 6) is similar between both datasets, validation loss (Fig. 7) strongly differs in favour of the clean dataset. Although part of this difference may come from the reduced number of audio files, another part is due to the better quality of the clean dataset. After training was completed, audio inferences of both networks were generated under the same conditions and compared manually. Subjectively, the evaluation indicated that the model trained using the clean dataset generated inferences with consistently less background noise and more stable stop token prediction, thus producing overall better results. Samples are provided on the dataset's website[10].

A further, automated analysis of 105 generated audio inferences from both models shows large differences with regard to minimum volume. While inferences generated by the clean model have an MSA of -57dB (MSA clean dataset -60dB), those produced by the model trained on the full dataset have -45dB (MSA full dataset -45dB). These discrepancies support the previously conducted, subjective evaluation and also prove the effect of applying this metric in the creation of clean datasets.

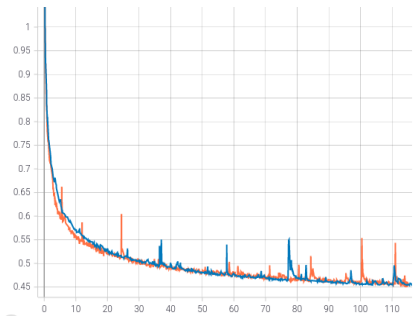
**Fig. 6.** Training Loss Tacotron 2 Hokuspokus orange: full, blue: clean

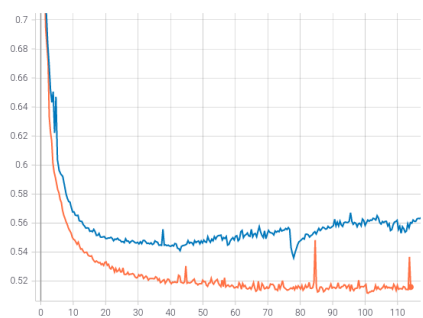
**Fig. 7.** Validation Loss Tacotron 2 Hokuspokus orange: full, blue: clean

## 5 Conclusion and Outlook

This paper describes the HUI-German-Audio-Corpus, a freely available, high-quality dataset for TTS in German consisting of audio transcript pairs of several speakers with a total length of over 300 hours. In addition, it contains a "clean" subset, which meets advanced quality criteria. While the audio to text alignment demonstrates a high degree of correctness, some manual steps such as normalization of ordinal numbers and abbreviations could be further assisted by a fitting deep neural network for POS tagging. We've demonstrated, that quality of the dataset is equally important as length. The higher frequency of 44.1 kHz compared to the 16 kHz of the MAILABs dataset makes a huge difference, although we've trained our samples only with 22.05 kHz. The fact, that we've normalized the text regarding numbers, which is especially demanding in German due to its different endings for numbers for different cases (e.g. genitive, dative), leads to good performance of the trained network models when reading numbers. Other datasets like the German part of MLS are completely missing the numbers and cannot be used for a TTS model that should be able to read numbers. The fact that Thorsten Müller's dataset is somewhat overemphasized leads to fast convergence and an easy to understand output, but an unnatural reading style. The large amount of data available for the voice Bernd Ungerer leads to a very stable output that can cope with almost any speech situation, despite its limited vocabulary used for training. The fact that our datasets contain both short and large audio parts leads to TTS models that are able to read longer texts in one piece.

---

[10] https://opendata.iisys.de/datasets.html#hui-audio-corpus-german

The usage of deep learning for the alignment of text and audio significantly increased the quality. TTS and automatic speech recognition (ASR) are closely related and can mutually benefit from each other. Training data for TTS can be reused for ASR as well. ASR can help to generate new training data for TTS. Sufficiently good TTS can on the other hand be used to generate additional training data for ASR, especially for words that are otherwise underrepresented in the training dataset. Generating statistics over the datasets and comparing the words present with manually curated lists like Wordnet can generate valuable insight for further tweaking. We envision to use TTS to generate audio for ASR that contains e.g. names, numbers and complex words in order to enhance existing audio datasets for ASR that are missing those. Finally, better text understanding can help splitting audio files at meaningful positions in the text, which would further enhance the already good training results.